\documentclass[aps,prb,twocolumn,superscriptaddress]{revtex4}
\usepackage{graphicx}
\usepackage{natbib}
\usepackage{color}
\begin{document}


\title{Nanoelectromechanical coupling in fullerene peapods probed via resonant electrical transport experiments}
\thanks{The final version of this preprint was published in \emph{Nature Communications} \textbf{1}, 37 (2010) doi: 10.1038/ncomms1034.}

\author{Pawel Utko}
\thanks{The first two authors have equally contributed to the present work. Correspondence and requests for materials should be addressed to P.U. or R.F. (pawel@fys.ku.dk or r.ferone@lancaster.ac.uk).}
\affiliation{Nano-Science Center \& Niels Bohr Institute, University of
Copenhagen, 
DK-2100 Copenhagen, Denmark}

\author{Raffaello Ferone}
\thanks{The first two authors have equally contributed to the present work. Correspondence and requests for materials should be addressed to P.U. or R.F. (pawel@fys.ku.dk or r.ferone@lancaster.ac.uk).}
\affiliation {Department of Physics, Lancaster University, Lancaster LA1 4YB, UK}
\affiliation{Department of Physics, University of Gothenburg, SE-412
96 G\"{o}teborg, Sweden}

\author{Ilya~V.~Krive}
\affiliation{Department of Physics, University of Gothenburg, SE-412
96 G\"{o}teborg, Sweden}
\affiliation{B.~I.~Verkin Institute for Low
Temperature Physics and Engineering, 
61103 Kharkov, Ukraine}

\author{Robert I.~Shekhter}
\affiliation{Department of Physics, University of Gothenburg, SE-412
96 G\"{o}teborg, Sweden}

\author{Mats Jonson}
\affiliation{Department of Physics, University of Gothenburg, SE-412
96 G\"{o}teborg, Sweden} 
\affiliation{SUPA, Department of Physics, Heriot-Watt University, Edinburgh EH14 4AS,
UK}
\affiliation{Division of Quantum Phases and Devices, School of Physics, Konkuk University, Seoul 143-701,
Korea}

\author{Marc Monthioux}%
\affiliation{
CEMES-CNRS, B.P. 94347, 
F-31055 Toulouse Cedex 4, France}%

\author{Laure No\'{e}}%
\affiliation{
CEMES-CNRS, B.P. 94347, 
F-31055 Toulouse Cedex 4, France}%

\author{Jesper Nyg{\aa}rd}
\affiliation{Nano-Science Center \& Niels Bohr Institute, University of
Copenhagen, 
DK-2100 Copenhagen, Denmark}


\begin{abstract}
Fullerene peapods, that is carbon nanotubes encapsulating fullerene molecules, can offer enhanced functionality with respect to empty nanotubes. However, the present incomplete understanding of how a nanotube is affected by entrapped fullerenes is an obstacle for peapods to reach their full potential in nanoscale electronic applications. Here, we investigate the effect of C$_{60}$ fullerenes on electron transport via peapod quantum dots. Compared to empty nanotubes, we find an abnormal temperature dependence of Coulomb blockade oscillations, indicating the presence of a nanoelectromechanical coupling between electronic states of the nanotube and mechanical vibrations of the fullerenes. This provides a method to detect the C$_{60}$ presence and to probe the interplay between electrical and mechanical excitations in peapods, which thus emerge as a new class of nanoelectromechanical systems.
\end{abstract}

\maketitle
Since their discovery \cite{Smith1998}, carbon nanotubes encapsulating arrays of C$_{60}$ fullerene molecules,\cite{Monthioux2002,Krive2006} the so-called fullerene peapods, have attracted a lot of interest among physicists, chemists, as well as materials and electrical engineers. Peapods are considered as potential building blocks in nanoelectronics, offering enhanced functionality with respect to empty nanotubes. Their prospective applications include data storage devices \cite{Kwon1999}, single electron transistors \cite{Yu2005,Utko2006,Quay2007,Eliasen2010}, and spin-qubit arrays for quantum computing \cite{Benjamin}. Nevertheless, it is still unclear to what extent the electronic properties of peapods, critical for their applications, are affected by the entrapped fullerenes.

Theoretical studies \cite{theory1, theory3, theory2} predict dramatic changes in the electronic band structure of the nanotube, due to hybridization and corresponding shifts of SWNT energy states and C$_{60}$ molecular orbitals. This is supported by scanning tunneling microscopy experiments \cite{Hornbaker2002} and photoluminescence excitation/emission mapping \cite{ Okazaki2008}. On the other hand, no fundamental perturbation of the nanotube electron system has been found in photoemission experiments \cite{Shiozawa2006} and electrical transport measurements on peapod quantum dots (QDs) \cite{Utko2006, Quay2007,Eliasen2010}. In the latter case, clean electron transport with signatures of Kondo physics \cite{Kondo1} has been observed via single or double QDs, in the regime of {\em strong} electronic coupling $\Gamma$ between energy states of the dot and its electrical leads. Coherent transport has also been found for peapods filled with Sc@C$_{82}$ metallofullerenes \cite{Cantone2008}, whose incorporation in principle should have an even stronger effect on the nanotube's band structure than C$_{60}$ molecules.

Here, we study the effect of the encapsulated C$_{60}$ fullerenes on electron transport via peapod quantum dots in the regime of {\em weak} electronic coupling. Compared to empty nanotubes, we find an abnormal temperature dependence of the Coulomb blockade (CB) conductance oscillations, indicating a modification of single-electron excitations at low temperatures. These novel elementary excitations, which appear as a result of C$_{60}$ encapsulation, can be explained by a nanoelectromechanical coupling between quantized electronic states of the nanotube and the mechanical vibrations of the fullerenes. Their appearance provides a method to detect the C$_{60}$ presence and to probe the interplay between electrical and mechanical degrees of freedom in fullerene peapods, which emerge as a new class of nanoelectromechanical systems (NEMS). We stress that other types of devices for which a coupling to mechanical vibrations plays a crucial role in electrical transport are molecular conductors \cite{Joachim2005} and transistors \cite{Luffe2008}, as well as suspended nanotubes \cite{Sapmaz2006, Leturcq2009}.

\section{Results}

\begin{figure} []
   \includegraphics[trim= 15mm 50mm 30mm 30mm, clip, width=8cm]{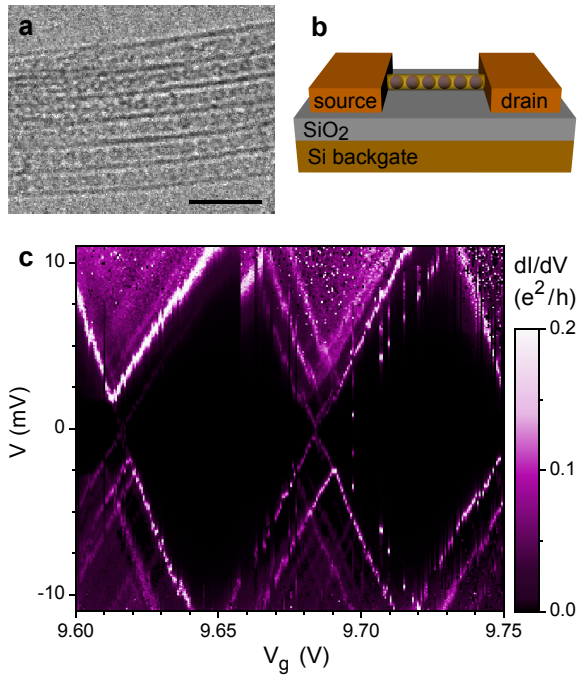}
     \caption{\textbf{C$_{60}$ fullerene peapod device.} (a) Transmission electron microscopy image of C$_{60}$ peapods with densely packed fullerene chains. The scale bar is 5~nm  long. (b) Schematic layout of the device. An individual C$_{60}$ peapod deposited on a SiO$_2$ substrate is contacted by Pd/Au source and drain electrodes, spaced 400 nm apart. (c) Stability diagram of a C$_{60}$ peapod. The (color coded) differential conductance $dI/dV$ is measured at $T=0.3$~K as a function of backgate voltage $V_g$ and bias voltage $V$, showing the diamond-shaped features characteristic for the Coulomb blockade.}
  \label{fig1}
 \end{figure}

Figure~\ref{fig1}a shows a transmission electron microscopy image of our peapod material, containing SWNTs densely filled with C$_{60}$ molecules. For electrical transport measurements, individual peapods with a diameter of 1.3-2~nm were contacted by Pd/Au source and drain electrodes, as schematically illustrated in Fig.~\ref{fig1}b. (See Methods for more details on devices.) The C$_{60}$ peapods showed clear Coulomb blockade behavior at low temperatures (Fig.~\ref{fig1}c), indicating that single-electron charging effects dominated the low-temperature electrical transport. Both the charging energy $U\approx8-12$~meV and the electronic level spacing $\Delta E \approx 2-3$~meV corresponded well to a 400~nm wide separation between the contact electrodes. Since the dots were probed at gate voltage ranges corresponding to a reduced electronic coupling $\Gamma$ to the leads, no zero-bias Kondo anomaly \cite{Utko2006, Quay2007} was observed here.

\begin{figure*} []
    \includegraphics[trim=10mm 133mm 15mm 15mm, clip, width=16.5cm]{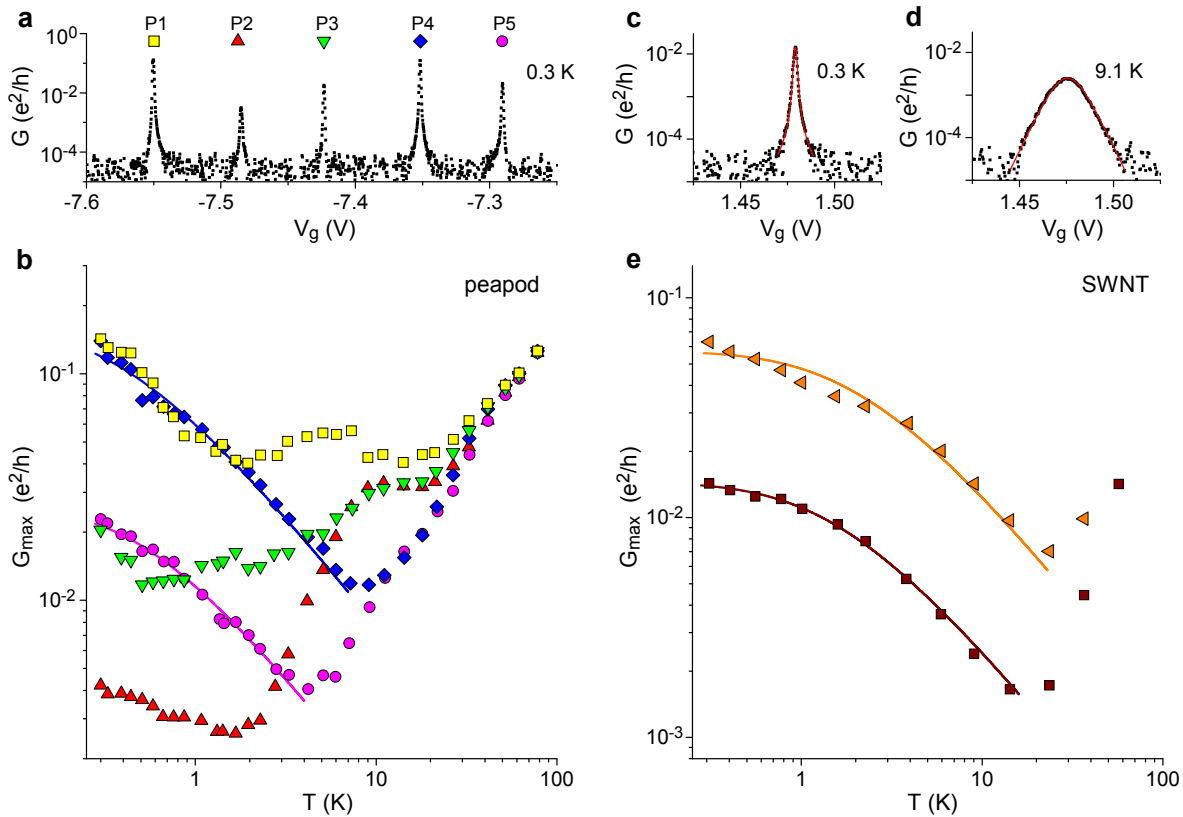}
     \caption{\textbf{Temperature dependence of linear conductance.} (a) Coulomb blockade oscillations in the linear conductance $G = dI/dV|_{V=0}$, measured as a function of gate voltage $V_g$ for a C$_{60}$ fullerene peapod. (b) Maximal conductance amplitude $G_\mathrm {max}$ vs temperature $T$, determined for the five peaks indicated in (a). Peaks P4 and P5 behave as predicted by the Breit-Wigner (BW) model,   while the others show an abnormal temperature dependence. (c) and (d) Lineshape of a conductance peak for a reference SWNT device, measured at (c) $0.3$~K, and (d) 9.1~K. These lineshapes correspond to a thermally broadened Breit-Wigner peak $G(E)=\int dE'(-\partial f(E')/\partial E') G_{\rm BW}(E-E')$. At low temperatures, $k_\mathrm BT \ll \Gamma$, the peak approaches the Lorentzian lineshape $G_{\rm BW}(E)\propto1/[(E-E_0)^2+(\Gamma/2)^2]$, as becomes apparent in (c). At higher temperatures, $\Gamma \ll k_\mathrm BT$, it can be approximated by a negative derivative of the Fermi function, $G(E)\propto-{\partial f}/{\partial E}$, see (d). (e) $G_\mathrm {max}$ versus $T$ for the reference SWNT device. As predicted by the BW theory, $G_\mathrm {max}(T)$ is nearly independent of $T$ at low temperatures, $k_\mathrm BT \ll \Gamma$, while at higher temperatures, $k_\mathrm BT \ll \Gamma$, the $1/T$ behavior is clearly recovered. In (b)-(e), the thin solid lines indicate fits to the experimental data performed using the Breit-Wigner model.}
     \label{fig2}
 \end{figure*}

Figure~\ref{fig2}a depicts the linear conductance $G$ of a fullerene peapod as a function of gate voltage $V_g$, revealing Coulomb blockade oscillations. This kind of experiment is a basic tool for probing both the energy spectrum and the wavefunctions of the elementary excitations in a quantum dot: The positions of the CB peaks provide information about the spectrum, while the magnitude of the tunneling current is proportional to the overlap between the wave functions of the incident electron and the elementary excitation in the dot. The temperature dependence of the heights of the five CB peaks that appear in Fig.~\ref{fig2}a is shown in Fig.~\ref{fig2}b.

The lineshape and temperature response of Coulomb blockade oscillations in the linear conductance has been widely investigated in the past \cite{Foxmann1994}, both for metallic and semiconductor quantum dots. In the CB regime where electron transport occurs only via a single electronic dot level, the tunneling probability as a function of energy is given by the  Breit-Wigner formula (see caption to Fig.~\ref{fig2}). The temperature dependence of the conductance peaks is therefore entirely due to thermal broadening of the energy of the incident electrons: The peak height is independent of temperature in the low-temperature limit, where $k_\mathrm B T\ll\Gamma$, while in the opposite high-temperature limit, $k_\mathrm B T\gg\Gamma$, a universal $1/T$-behavior appears. This corresponds to what we observed for empty SWNTs, as shown in Figs.~\ref{fig2}c-e.

\begin{figure*} []
 \includegraphics[trim=10mm 115mm 10mm 16mm, clip, width=16.5cm]{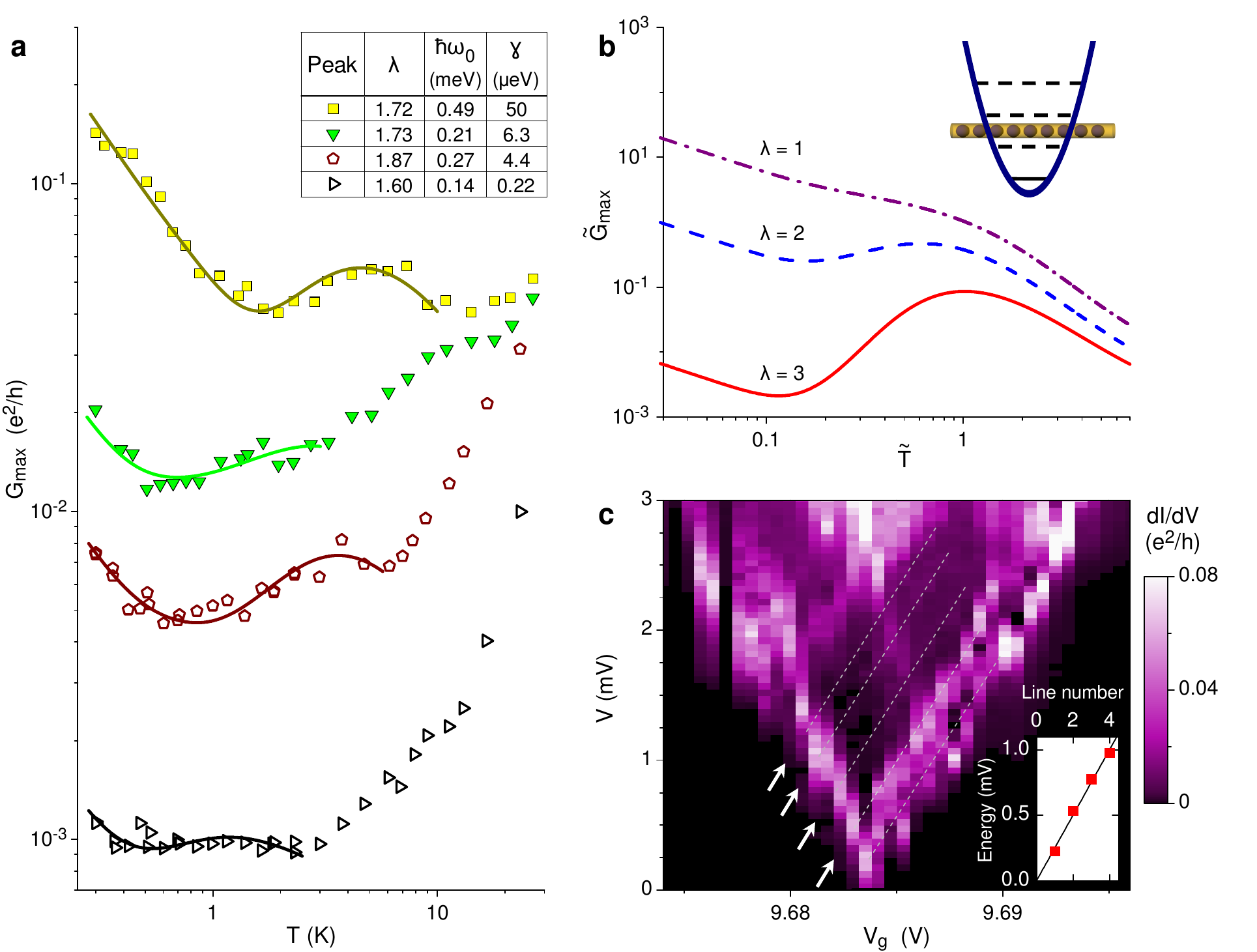}
    \caption{\textbf{Polaronic interactions in a fullerene peapod.} (a) $G_\mathrm {max}$ vs $T$ for conductance peaks with an abnormal temperature behavior with respect to the Breit-Wigner model. Solid and open symbols indicate the data points measured for two different peapod devices. The solid lines are fits to the experimental data, calculated using the polaronic model described in the text. The inset gathers the determined values of the fitting parameters. (b, Inset) A schematic diagram of the vibronic confining  potential. The black parabola represents the harmonic potential defining the oscillating motion of the vibrator, while horizontal lines are vibronic energy states spaced by $\hbar\omega_0$. At temperatures $k_\mathrm BT\ll \hbar\omega_0$, only the vibronic ground state (solid line) participates in the conductance. As soon as $k_\mathrm BT$ becomes of the order of $\hbar\omega_0$, thermally excited vibronic modes of the oscillating scatterer are generated (dashed lines) with characteristic energy $n(z)\hbar\omega_0$, where $n(z)$ is the Bose distribution with $z=\hbar\omega_0 / k_\mathrm B T$. The ratio between this energy and the polaronic energy shift $\lambda^2\hbar\omega_0$ determines how detrimental the thermal vibrations can be to the polaronic effects. (b, Main figure) Calculated renormalized peak conductance $\widetilde{G}_\mathrm {max}=G_\mathrm {max}\hbar\omega_0/\gamma$ as a function of reduced temperature $\widetilde{T}=k_\mathrm BT/\hbar\omega_0$ for three different values of the polaronic coupling constant $\lambda$. At low temperatures, $k_\mathrm BT\ll\hbar\omega_0$, an exponential suppression of the conductance occurs, $G_\mathrm {max}\propto G_{\mathrm{max},\lambda=0}(T) \exp(-\lambda^2)$, due to a reduced overlap between the electronic wave function in the lead and the polaronic wave function in the dot. Here, $G_{\mathrm {max},\lambda=0}(T)$ is the thermally broadened BW peak conductance for noninteracting electrons. The stronger the polaronic coupling $\lambda$, the stronger the suppression of the conductance is. By increasing the temperature,  the competition between the thermally generated vibrons and the polarons gives rise to a non-monotonic behavior. An abnormal temperature dependence of the peak height is particularly well visible for $k_\mathrm B T\lesssim \lambda^2\hbar\omega_0$ where the conductance is dominated by the term $(1/\sqrt{T})\exp (-\lambda^2\hbar\omega_0/4k_\mathrm B T)$. At higher temperatures, $k_\mathrm BT\gg\lambda^2\hbar\omega_0$, the polaronic blockade is completely overcome by thermal excitations and the conductance scales as $1/T$, following the Breit-Wigner formula. See Methods section for more details on the model. (c) Closer look at a stability diagram $dI/dV$ vs $V$ vs $V_g$. Arrows indicate weak quasi-periodic excitation lines running parallel to the Coulomb diamond edge. The inset shows their excitation energies; the solid line has a slope of 0.25~meV.}
  \label{fig3}
\end{figure*}

For fullerene peapods, however, the measured temperature dependence of the conductance peaks is often significantly different (see Fig.~2b), implying that the Breit-Wigner picture is no longer valid. We interpret this discrepancy as a signature of a change in the spectrum of the elementary excitations and of their wavefunctions, a change that can be associated with the encapsulated fullerenes. Typically, as highlighted in Fig.~\ref{fig3}a, an anomalous temperature dependence appears on the scale of a few kelvin. The corresponding energy is of the order of $0.1$ meV, which is small on the electronic energy scale but relevant for the longitudinal vibrational motion of the encapsulated fullerenes. (For the low-diameter peapods used in our study, the characteristic scale of transverse vibrations would be $\sim$1 meV.)

In order to interpret our experimental results, we use a model where SWNT electrons are electromechanically coupled to a set of low-frequency vibrators: single C$_{60}$ fullerenes \cite{Krive2008} or their clusters trapped by a longitudinal harmonic potential inside the nanotube. Within this approach, the peapod is treated as a multilevel quantum dot for which each quantized electronic level is selectively coupled to a single vibrational mode, in a similar manner as in the local Holstein model \cite{Galperin2007}, and the corresponding dimensionless electron-vibron coupling constant $\lambda$ is proportional to the amplitude of the vibronic quantum fluctuations. As a result, the resonant tunneling electrons can be dressed by vibrational excitations if they spend sufficiently long time in the dot to form real electron-vibron hybrids -- Holstein polarons \cite{Krive2008}. Since the probability of polaron formation is enhanced for long electron dwell times in the quantum dot, $1/\Gamma$, polaronic effects become more pronounced in the regime of weak electronic coupling $\Gamma$.

We stress that the electron-vibron coupling does not have any significant effect on the fullerene vibrations themselves, as they are mostly determined by the C$_{60}$ confinement within the carbon nanotube. Even though the coupling to each single electronic level may be strong and depend on the position of the oscillating fullerenes, their total energy shift averages out over a large number of individual electronic states in the dot.

Our model calculations (Fig.~\ref{fig3}b) have revealed qualitatively similar features to those experimentally observed for conductance peaks deviating from conventional Breit-Wigner behavior (Fig.~\ref{fig3}a). In particular, three characteristic thermal regimes can be distinguished:
(i) When the thermal energy is much smaller than the vibration energy, $k_\mathrm BT\ll\hbar\omega_0$, the conductance becomes suppressed for strong electron-vibron coupling $\lambda \gtrsim 1$. Such a \emph{polaronic} (or Franck-Condon) blockade of conductance \cite{Koch2005} signals the reduced probability for an electron to tunnel from a pure electronic state in the lead to a polaronic state in the dot, and vice versa. (ii) As the temperature rises, the population of the thermally excited vibronic levels increases which has a destructive effect on the polaronic wavefunction. This leads to a non-monotonic behavior of the conductance in the temperature interval $k_\mathrm BT\lesssim\lambda^2\hbar\omega_0$, where $\lambda^2\hbar\omega_0$ represents the characteristic energy of a polaron. (iii) At high temperatures, $k_\mathrm BT\gg\lambda^2\hbar\omega_0$, polaronic effects are lifted and  the conductance takes standard Breit-Wigner form $G_\mathrm {max} \propto 1/T$.

To determine the electron-vibron coupling $\lambda$ in the peapod devices, we have numerically fitted the temperature dependence of conductance maxima (Fig.~\ref{fig3}a) with the thermal progression predicted by our model (Eq.~\ref{GT2} in Methods). The obtained fits are in good agreement with the experimental data, yielding large values for the polaronic coupling parameter, $\lambda=1.6-1.9$. These values are consistent with the suppression of the conductance observed at low temperatures. They are also of the same order of magnitude as in previous experiments on suspended carbon nanotube devices \cite{Sapmaz2006, Leturcq2009} where the vibronic states were due to intrinsic modes of the SWNT. (Note that in Refs.~\onlinecite{Sapmaz2006} and \onlinecite{Leturcq2009} the coupling constant was defined as $g=\lambda^2$). The inset to Fig.~\ref{fig3}a gathers all the fitting parameters extracted for the investigated peaks.

The vibrational origin of the abnormal features observed in the  $G_\mathrm {max}(T)$ is in addition supported by the appearance of weak, quasi-periodic excitation lines along the edges of the Coulomb blockade diamonds in the stability diagrams, as indicated by arrows in Fig.~\ref{fig3}c. The mean energy spacing $\sim 0.25$~meV of these nearly equidistant excitations is an order of magnitude smaller than the electronic level spacing $\Delta E = 2-3$~meV in our 400~nm long peapod quantum dots. On the other hand, it closely matches the values of vibrational quanta extracted from the fits to $G_\mathrm{max}(T)$, see the inset to Fig.~\ref{fig3}a.

\section{Discussion}
As mentioned before, our experimental data clearly indicate that not all of the studied conductance peaks show abnormal features in their temperature dependence, see Fig.~\ref{fig2}b. This observation can be explained if, for different SWNT electron states $n$, there are strong fluctuations \cite{Krive2008} in the value of the polaronic coupling constant. Such fluctuations may be due to a local character of the electromechanical coupling. Consequently, only for some electronic levels the characteristic time needed to form a polaronic state $1/(\lambda_n^2 \hbar \omega_0)$ is sufficiently shorter than the electron dwell time in the dot $1/\Gamma_n$.

We note here that the electronic coupling $\Gamma$ could not be directly determined from the polaronic fits. Instead, we were able to probe a related parameter $\gamma=\Gamma_L\Gamma_R/(\Gamma_L+\Gamma_R)$, yielding $\gamma=0.004-0.050$~meV for the investigated conductance peaks. Here, $\Gamma_L$ and $\Gamma_R$ are electronic couplings to the left and to the right reservoir of the quantum dot, respectively, and $\Gamma=\Gamma_L+\Gamma_R$. Assuming symmetric barriers, $\Gamma_L=\Gamma_R$, the value of $\Gamma$ could thus be evaluated as $\Gamma=4\gamma$. For the asymmetric coupling, $\Gamma_L=10\Gamma_R$, this would yield $\Gamma=12.1\gamma$.

In order  to gain some insight into the nature of the vibrational motion of fullerene molecules trapped inside  a SWNT,  we have performed a simple estimation of the potential that confines their longitudinal motion. The confining potential can readily be obtained from Girifalco's universal potential for graphitic structures  \cite{Girifalco} if we assume that it is mainly due to the interaction of a vibrating C$_{60}$ molecule with its nearest-neighbor molecules (one on each side). The resulting potential is well approximated by a harmonic potential characterized by the force constant $k=2.2$~Nm$^{-1}$. The corresponding vibrational energy quantum $\hbar\omega_0$ matches the value of about $0.25$~meV, needed to fit the experiments, for a particle mass about ten times larger than that of a single C$_{60}$ molecule ($M_{\rm C_{60}}=1.2\cdot10^{-24}$~kg). The implication is that the relevant molecular vibrations may involve clusters of about ten C$_{60}$s. For peapods with densely packed fullerene chains, clusters of this size could form inside the SWNT, for example, as a result of local tube corrugations or presence of defects (openings) on SWNT side wall \cite{Monthioux2002}.

The present incomplete understanding of how a nanotube is affected by encapsulated fullerenes is an obstacle for peapods to reach their full potential in nanoelectronic applications \cite{Yu2005,Utko2006,Quay2007,Eliasen2010,Kwon1999,Benjamin}. The experiments presented here shed light on this issue: They demonstrate the presence of nanoelectromechanical coupling in carbon peapods which thus emerge as a new class of nanoelectromechanical systems. Moreover, the characterization technique described here can be thought of as a tool for detecting the presence of fullerenes, or other molecules, in a nanotube. Such an experimental verification can be performed \emph{in situ} for individual tubes in electronic devices, unlike in other detection techniques which require either a large ensemble of peapods (Raman spectroscopy \cite{Kuzmany2004}, x-ray \cite{Launois2010}) or probing in dedicated systems (transmission electron microscopy, scanning tunneling microscopy \cite{Hornbaker2002}, low-temperature atomic force microscopy \cite{Ashino2008}).

Further studies using suspended peapods could allow a tuning of the fullerenes' spatial distribution within the nanotube by means of mechanical bending deformations of the SWNT. If the resonant character of electron transmission could be controlled mechanically by these means or others, carbon peapods could find applications as NEMS resonators with an operation frequency set by the vibrational motion of the encapsulated fullerenes in the 100 GHz range.

\section{Methods}
\subsection{Devices}
The high-quality fullerene peapods investigated in our study were based on arc discharge-grown SWNTs. The empty tubes (also used as the reference SWNTs) were supplied by NANOCARBLAB as a purified material (80\% pure grade). Due to the purification procedure (multi-step heating in air, soaking in HNO$_3$, microfiltration), the as-received SWNTs were already opened and required no further treatment prior to their filling with C$_{60}$ fullerenes (98\% purity, INTERCHIM). The filling was carried out via the vapor phase method \cite{Monthioux2002}, in an evacuated and sealed ampoule at $500 ^\circ$C for 24 hours. In order to remove excess fullerenes, heat treatment was afterwards performed under dynamic vacuum, at $800 ^\circ$C for 1 hour. Transmission electron microscopy indicated that 90-95\% of the tubes in the resulting material encapsulated fullerenes (i.e., were peapods). A quantitative analysis by x-ray diffraction \cite{Launois2010} confirmed that the peapods were densely packed with C$_{60}$s, with a filling ratio of $\sim80$\%. In addition, signatures of the encapsulated C$_{60}$ fullerenes were found in Raman spectroscopy.

The peapods (or the reference empty nanotubes) were sonicated in dichloroethane and deposited on a silicone wafer covered with silicone oxide. Individual tubes with a diameter of $1.3-2$\,nm were located and selected for further processing by means of atomic force microscopy. Electron beam lithography was then used to define source and drain electrodes of Pd/Au (15\,nm/45\,nm), with a spacing of 400\,nm. A highly n-doped Si substrate, situated beneath the 500\,nm-thick SiO$_2$ layer, served as a backgate. Electrical transport experiments were performed in a $^3$He cryostat with a base temperature of $T=300$~mK.

\subsection{Model}
We highlight here the main steps of our theoretical modeling, which is based on the local Holstein model\cite{Galperin2007}. More details can be found in Ref. [\onlinecite{Krive2008}].

In order to get analytical results, we assume that only one vibrating scatterer (a cluster of fullerenes or a single molecule) is confined inside the SWNT.
The full Hamiltonian for a two-terminal peapod QD system reads:
\begin{equation}\label{hamilt}
{\cal H}= H_\mathrm{L}+H_\mathrm{R}+H_{\mathrm T,\mathrm{L}}+H_{\mathrm T,\mathrm{R}}+H_\mathrm{QD} ,
\end{equation}
where $H_\alpha=\sum_{k_\alpha}\varepsilon_{k_\alpha} c^\dagger_{k_\alpha} c_{k_\alpha}$ is the Hamiltonian for the Fermi leads and $\alpha={\mathrm{L,R}}$ for the left and right reservoir, respectively. The peapod-QD Hamiltonian reads:
\begin{equation} \label{Ham}
H_\mathrm{QD}=\sum_n\varepsilon_nc_n^{\dagger}c_n
+\sum_nV_n(L,l)c_n^{\dagger}c_n(b^{\dagger}+b)
+\hbar\omega_0b^{\dagger}b ,
\end{equation}
where $\{\varepsilon_n\}$ is the set of energy levels in the dot,
$\hbar\omega_0$ is the vibrational quantum of energy, $V_n(L,l)=\hbar
v_Fx_0(\partial k_n/\partial x)$, and $x_0$ is the
amplitude of the zero-point fluctuations of the bosonic field. $c_n (c_n^{\dagger})$ and $b (b^{\dagger})$ are fermionic and bosonic operators with canonical commutation relations. The longitudinal motion of the scatterer in the SWNT is defined by a harmonic potential. Finally, the tunneling Hamiltonian reads:
\begin{equation}
H_{\mathrm T,\alpha}=\sum_{k_\alpha,n}t\,(c^\dagger_{k_\alpha} c_n+\mathrm{h.c.})\;,\label{tunham}
\end{equation}
where we suppose that the hopping matrix elements $t$ between the two leads and the dot are of the same order.

The above QD Hamiltonian can be diagonalized by a unitary transformation, which reveals that the major effect of the vibrational bosonic field is to renormalize the energy levels in the dot. These levels are shifted by an amount $\lambda_n^2 \hbar\omega_0$, the so called polaronic shift, where $\lambda_n=V_n(L,l)/(\hbar\omega_0)$ is the electron-vibron coupling constant. To obtain the expression for the conductance of our system, we start from the Meir-Wingreen formula \cite{Meir}

\begin{equation}
I(V)=-\frac{2 e}{h}\int d\varepsilon \,[f_\mathrm{L}(\varepsilon)-f_\mathrm{R}(\varepsilon)]\Im \mathrm m\{\mathrm{Tr}[\gamma\,
G_\mathrm{QD}^\mathrm{ret}(\varepsilon)]\},\label{curr}
\end{equation}
where $f_\mathrm{L/R}(\varepsilon)$ are the Fermi functions for the left and right reservoirs,
$\gamma=\Gamma_\mathrm{L}\Gamma_\mathrm{R}/(\Gamma_\mathrm{L}+\Gamma_\mathrm{R})$;
$\Gamma_\mathrm{L/R}=2\pi t^2 \rho_\mathrm{L/R}$ are the widths of the resonance due to tunneling from the left
or right lead, and $\rho_\mathrm{L/R}$ are the densities of states in the reservoirs. $G_\mathrm{QD}^\mathrm{ret}(\varepsilon)$ is the retarded Green function of the dot which can be evaluated in the wide-band limit\cite{Wingreen}.

By means of the gate voltage, an energy level in the QD can be tuned to line up with the Fermi level $\varepsilon_\mathrm F$ in the leads, so that $\varepsilon_\mathrm F=\varepsilon_n-\lambda_n^2 \hbar\omega_0$. This is when the conductance reaches its maximum. In the linear conductance regime, one finds that the temperature dependence of these maxima is given by the simple expression
\begin{equation} \label{GT2}
G_{\mathrm {max},\lambda}(T)=G_{\mathrm {max},\lambda=0}(T)F_{\lambda}(z),
\end{equation}
where
\begin{equation} \label{GT3}
  G_{\mathrm {max},\lambda=0}(T)\simeq \frac{\pi}{2} \frac{e^2}{h}\frac{\gamma}{k_\mathrm BT}
\end{equation}
is the standard resonance conductance of a single-level quantum dot. In equation (\ref {GT2}), the  function $F_{\lambda}(z)$, with $z=\hbar\omega_0/k_\mathrm BT$, can be expressed as a series,
\begin{eqnarray} \label{Fx}
F_{\lambda}(z)&=&\exp\{-\lambda^2[1+2n(z)]\}\nonumber\\
&\times&\sum_{l=-\infty}^{\infty}\frac{\exp(-lz/2)I_l[2\lambda^2\sqrt{n(z)(1+n(z))}]}
{\cosh^2(lz/2)},    
\end{eqnarray}
where $I_l$ is the modified Bessel function of the first kind and $n(z)$ is the Bose distribution function. The thin solid lines in Fig.~\ref{fig3}a show fits to the experimental data obtained with the help of expression (\ref{GT2}) for the conductance.

\section{Acknowledgements}
R.F. thanks A. Isacsson, L. Oroszl\'any, R. Rehammar for fruitful discussions. Financial support from EC FP6 funding (Contract No. FP6-2004-IST-003673), the Swedish VR and SSF, the CLIPS project of the Danish Strategic Research Council, and the Korean WCU programme funded by MEST through NFR (R31-2008-000-10057-0) is gratefully acknowledged. R.F. gratefully acknowledges financial support from EPSRC First Grant EP/E063519/1.

\section{Authors contributions}
R.I.S., P.U., R.F., I.V.K. and J.N. planned the experiment. L.N. and M.M. fabricated peapod material. P.U. fabricated the samples and carried out the measurements. P.U., R.F., I.V.K., J.N. and R.I.S. analyzed the data. R.F., I.V.K., M.J. and R.I.S. did the theoretical modeling. R.F. carried out the theoretical simulations. P.U. and  R.F. wrote the paper, with inputs from M.J., R.I.S., and J.N.

\end{document}